\documentclass[useAMS,usenatbib]{mn2e}

\usepackage{amssymb}
\usepackage{amsfonts}
\usepackage{mncite}
\usepackage{epsfig}
\usepackage{psfig}

\defcitealias{LR04}{Paper I}

\begin{document}

\title[Investigating fragmentation in discs]{Investigating fragmentation
conditions in self-gravitating accretion discs}

\author[W.K.M. Rice, G. Lodato, \& P.J. Armitage]{W.K.M. Rice$^1$,
G. Lodato$^2$ and P.J. Armitage$^{3,4}$ \\ 
$^1$ Institute of Geophysics and Planetary Physics, 
University of California, Riverside, CA 92521, USA \\
$^2$ Institute of Astronomy, Madingley Road, Cambridge, CB3 0HA
\\ $^3$ JILA, Campus Box 440, University of Colorado, Boulder, CO
80309-0440, USA
\\ $^4$ Department of Astrophysical and Planetary Sciences, University
of Colorado, Boulder, CO 80309-0391, USA}
 
\maketitle

\begin{abstract}
The issue of fragmentation in self-gravitating gaseous accretion
discs has implications both for the formation of stars in discs in the
nuclei of active galaxies, and for the formation of gaseous planets or
brown dwarfs in circumstellar discs. It is now well established that
fragmentation occurs if the disc is cooled on a timescale smaller than
the local dynamical timescale, while for longer cooling times the disc
reaches a quasi-steady state in thermal equilibrium, with the cooling
rate balanced by the heating due to gravitational stresses. We
investigate here how the fragmentation boundary depends on the assumed
equation of state. We find that the cooling time required for
fragmentation increases as the specific heat ratio $\gamma$ decreases,
exceeding the local dynamical timescale for $\gamma = 7/5$. This result
can be easily interpreted as a consequence of there being a maximum
stress (in units of the local disc pressure) that can be sustained by a
self-gravitating disc in quasi-equilibrium. Fragmentation occurs if the
cooling time is such that the stress required to reach thermal
equilibrium exceeds this value, independent of $\gamma$. This result
suggest that a quasi-steady, self-gravitating disc can never produce a
stress that results in the viscous $\alpha$ parameter exceeding $\sim
0.06$.
\end{abstract}


\section{Introduction}
It is becoming clearer that self-gravity may play an important role in
the dynamical evolution of accretion discs. Active galactic nuclei
(AGN) often show rotation curves that depart significantly from a
Keplerian profile \citep{greenhill97,LB03a,kondratko05}, while
protostellar disc masses may be a significant fraction of the central
object mass during the early stages of star formation
\citep{linpringle87,larson84}. For example, a recent radio observation
of a Class 0 object (a very young protostellar source, with age $<
10^5$ yrs and mass $M_{\star}\approx 0.8M_{\odot}$,
\citealt{rodriguez05}) shows a disc with a estimated mass $M_{\rm
disc}=0.3-0.4M_{\odot}$ and a very suggestive two armed spiral-like
stucture. In addition, there are also clues that massive accretion
discs can be present around massive protostars
\citep{beltran04,chini04}.

An important effect of disc self-gravity is that it provides an
efficient mechanism for transporting angular momentum outwards,
allowing mass to accrete onto the central object
\citep{linpringle87,laughlin94}. In protostellar discs, in
which the ionization level is expected to be low, thus inhibiting
MHD-driven turbulence \citep{matsumoto95,gammie96}, this may be the dominant
mechanism, at least in certain regions of the disc. In general,
however, the transport associated with self-gravitating disturbances
may not be well described as a simple diffusion mechanism.
\citet{balbus99} have shown that, for self-gravitating discs, the
energy flux contains non-local terms that they associate with wave
energy transport. On the other hand, \citet{LR04,LR05} have shown that,
for a self-gravitating accretion disc in thermal equilibrium, the
dissipation arising from gravitational stresses agrees reasonably well
with the expectations based on the standard viscous theory
\citep{shakura73, pringle81} (see, however, \citealt{mejia05} for a
different result).

In the simulations performed by \citet{LR04,LR05} thermal equilibrium
is achieved by allowing the disc to heat up through $P\mbox{d}V$ work
and shock dissipation, and cooling it at a prescribed rate. They use a
simple cooling term with a cooling time given by $t_{\rm cool} = \beta
\Omega^{-1}$, where $\Omega$ is the angular frequency and $\beta =
7.5$. In thermal equilibrium it can be shown \citep{pringle81,gammie01}
that, for a viscous disc, the parameter $\alpha$ \citep{shakura73},
that characterises angular momentum transport, and the cooling time,
$t_{\rm cool}$, are related through
\begin{equation}
\alpha=\frac{4}{9 \gamma (\gamma - 1)}\frac{1}{t_{\rm cool} \Omega},
\label{alpha}
\end{equation}
where $\gamma$ is the ratio of the specific heats. 

The ultimate evolution of a self-gravitating accretion disc depends
strongly on the rate at which the disc heats up, through the growth of
the instability, and on the rate at which it cools.  It has been shown,
using a local, two-dimensional model \citep{gammie01}, that if $t_{\rm
cool} < 3 \Omega^{-1}$, the disc will fragment into bound objects
rather than evolve into a quasi-steady state, a result that was largely
confirmed by \citet{rice03c} using global, three-dimensional models.
This process has been suggested as a mechanism for forming both gaseous
planets in protostellar discs \citep{kuiper51,boss98,mayer04}, and for forming
stars in active galactic discs \citep{goodman04}.

A major obstacle to the formation of objects via disc fragmentation is
the requirement that the cooling time be smaller than the local
dynamical time \citep{gammie01,rice03c}.  This may be possible in AGN
disc \citep{johnson03}, but appears unlikely in protostellar discs
\citep{rafikov05}. The cooling time requirement was, however,
determined using equations of state with specific heat ratios of
$\gamma = 2$ \citep{gammie01}, and $\gamma = 5/3$ \citep{rice03c}.  It
has been suggested \citep{LR05} that since the stress required to
balance the cooling rate (as measured by the viscous $\alpha$) depends
on the specific heat ratio, $\gamma$ (see equation \ref{alpha}), the
cooling time required for fragmentation may also depend on $\gamma$.
In particular, if there is a maximum stress $\alpha_{\rm max}$ (in
units of the local disc pressure) sustainable by a self-gravitating
disc, then equation (\ref{alpha}) states that fragmentation should occur
if:

\begin{equation}
\label{eq:amax}
t_{\rm cool}\Omega<\frac{4}{9\gamma(\gamma-1)}\frac{1}{\alpha_{\rm max}}.
\end{equation}

In this paper we investigate, using global, three-dimensional
simulations, how the fragmentation boundary (as measured by $t_{\rm
cool}$) varies for different values of $\gamma$. We also consider
various disc masses to study the suggestion by \citet{rice03c} that the
fragmentation boundary may depend on the ratio of the disc mass to the
mass of the central object. In \S2, we describe the range of cooling
times, disc masses, and specific heat ratios, $\gamma$,  that have been
considered, and determine, for a given disc mass, the cooling time
required for fragmentation. In \S3 we discuss these results in light of
the relationship between the stresses in the disc (as measured by the
viscous $\alpha$) and the imposed cooling time. We show that
fragmentation is indeed easier in discs with smaller specific heat
ratios. We therefore conclude that there is a maximum stress
sustainable by a self-gravitating disc and we quantify this maximum
stress to be $\alpha_{\rm max}\approx 0.06$. In \S4 we discuss our
results and draw our conclusions.

\section{Simulation results}
\label{simres}

The simulations performed here are very similar to those of
\citet{rice03c} and \citet{LR04,LR05}. The three-dimensional, gaseous
discs are modelled using Smoothed Particle Hydrodynamics (SPH)
\citep{monaghan92,benz90}, a Lagrangian hydrodynamics code. The disc
is represented by 250000 SPH particles, while the central star is a
point mass onto which gas particles may accrete if they approach to
within the accretion radius (here taken to be at a radius of $R_{\rm
  acc} = 0.25$). In code units, the disc extends from $R_{\rm in} = 0.25$
to $R_{\rm out} = 25$, and the central object has a mass of $M_* = 1$.
We consider disc masses of $M_{\rm disc} = 0.1, 0.25$ and $0.5$, with
initial surface density profiles of $\Sigma \propto r^{-1}$, and with
a temperature that has a radial profile of $T \propto r^{-0.5}$. With
these initial surface density and temperature profiles, the Toomre
stability parameter, $Q = c_s \Omega / \pi G \Sigma$, is not initially
constant, but decreases with increasing radius.  The temperature is
therefore normalised such that at the beginning of the simulation the
disc is stable, with a minimum $Q = 2$ at $R = 25$.

Since we are interested in how cooling influences the disc evolution,
the disc gas is allowed to heat up due to both $P{\rm dV}$ work and
viscous dissipation, with the viscosity given by the standard SPH
artificial viscosity \citep{monaghan92} with $\alpha_{\rm SPH} =
0.1$, and $\beta_{\rm SPH} = 0.2$.  We use an adiabatic equation
of state and consider specific heat ratios of $\gamma = 5/3$, and
$\gamma = 7/5$. A particle with internal energy per unit mass $u_i$ is
then cooled using
\begin{equation}
\frac{{\rm d} u_i}{{\rm d} t} = - \frac{u_i}{t_{\rm cool}}
\end{equation}
where $t_{\rm cool} = \beta \Omega^{-1}$. 

An important numerical issue to be considered in this context is the
role of artificial SPH viscosity. The growth and the saturation of
gravitational instabilties depends on the balance between external
cooling and internal heating provided by the instability
itself. Therefore, we have to be sure that dissipation is dominated by
gravitational instabilities rather than by artificial viscosity (which
would provide an extra, undesired, stabilization term in the energy
balance). It can be shown \citep{lubow94,murray96} that artificial SPH
viscosity scales as $\nu_{\rm SPH}\propto \alpha_{\rm SPH}\langle
h\rangle$, where $\langle h\rangle$ is the average SPH smoothing
length. We therefore should achieve a high resolution (in order to keep
the smoothing length small) and adopt a sufficiently low value of
$\alpha_{\rm SPH}$, while preserving the ability of the code to
properly resolve the shocks that arise in the simulation. We have
already shown (\citealt{LR04}, Appendix) that with the setup described
above ($N=250000$, $\alpha_{\rm SPH}=0.1$ and $\beta_{\rm SPH} = 0.2$),
we are indeed able to properly resolve the shocks and to have an
artificial dissipation smaller by more than one order of magnitude with
respect to gravitationally induced dissipation. We are therefore
confident that artificial viscosity is not going to affect
significantly our conlcusions.

For each disc mass, and for each $\gamma$, we have performed a large
number of simulations with different values of $\beta$. We initially
start with a $\beta$ value that should result in fragmentation
\citep{gammie01,rice03c}. We stop the simulation once at least one
clump/fragment has formed that has a density $2-3$ orders of magnitude
greater than the surrounding gas.  The densest clump is then tested to
check if it is bound. Firstly, we determine the approximate size of the
clump, by finding the distance from the center of the clump at which
the density has returned to a value comparable with that of the
surrounding disk. All the particles within this spherical volume are
then assumed to be part of this clump.  In every case in which
fragmentation occurred, the densest clump consisted of at least 100
particles, and in some cases as many as 500 particles.  This more than
satisfies the Jeans criterion \citep{bate97}, and we are therefore
confident that the fragmentation in these simulations is not
artificial.  Once the clump size has been determined, we then calculate
the total thermal energy and the gravitational potential energy.  If
the net energy is negative the clump is bound, the simulation is
stopped, and a new simulation is started with the same initial
conditions, but with $\beta_{\rm new} = \beta_{\rm old} + 1$.  This new
simulation is then run until it either also fragments or until the disk
has evolved into a quasi-steady state \citep{gammie01,rice03c} without
any signs of fragmentation. We also run the non-fragmenting simulations
approximately an outer rotation period longer than the equivalent
simulation that did undergo fragmentation.

\begin{figure}
\centerline{\epsfig{figure=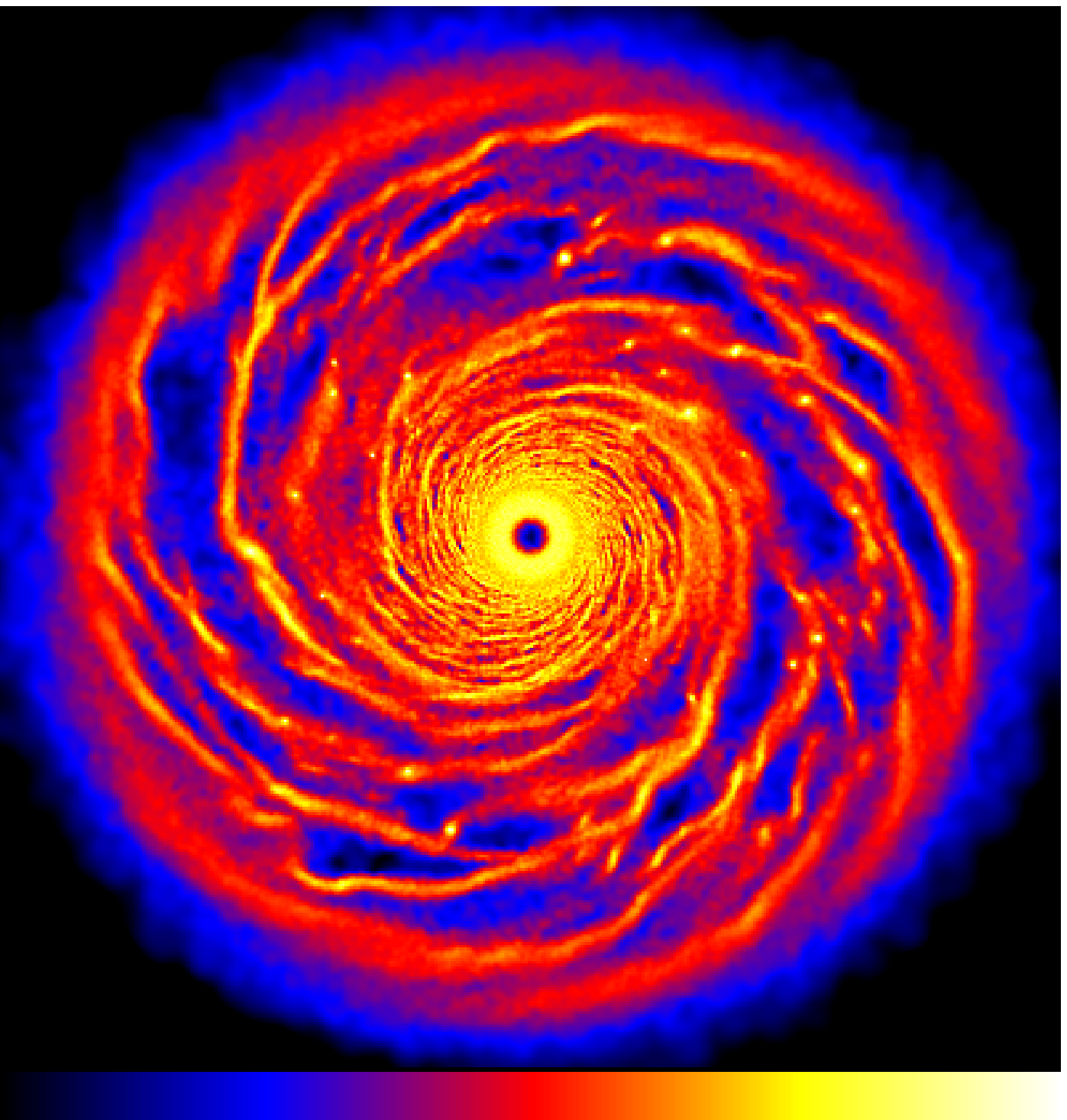,width=40.0mm}
            \epsfig{figure=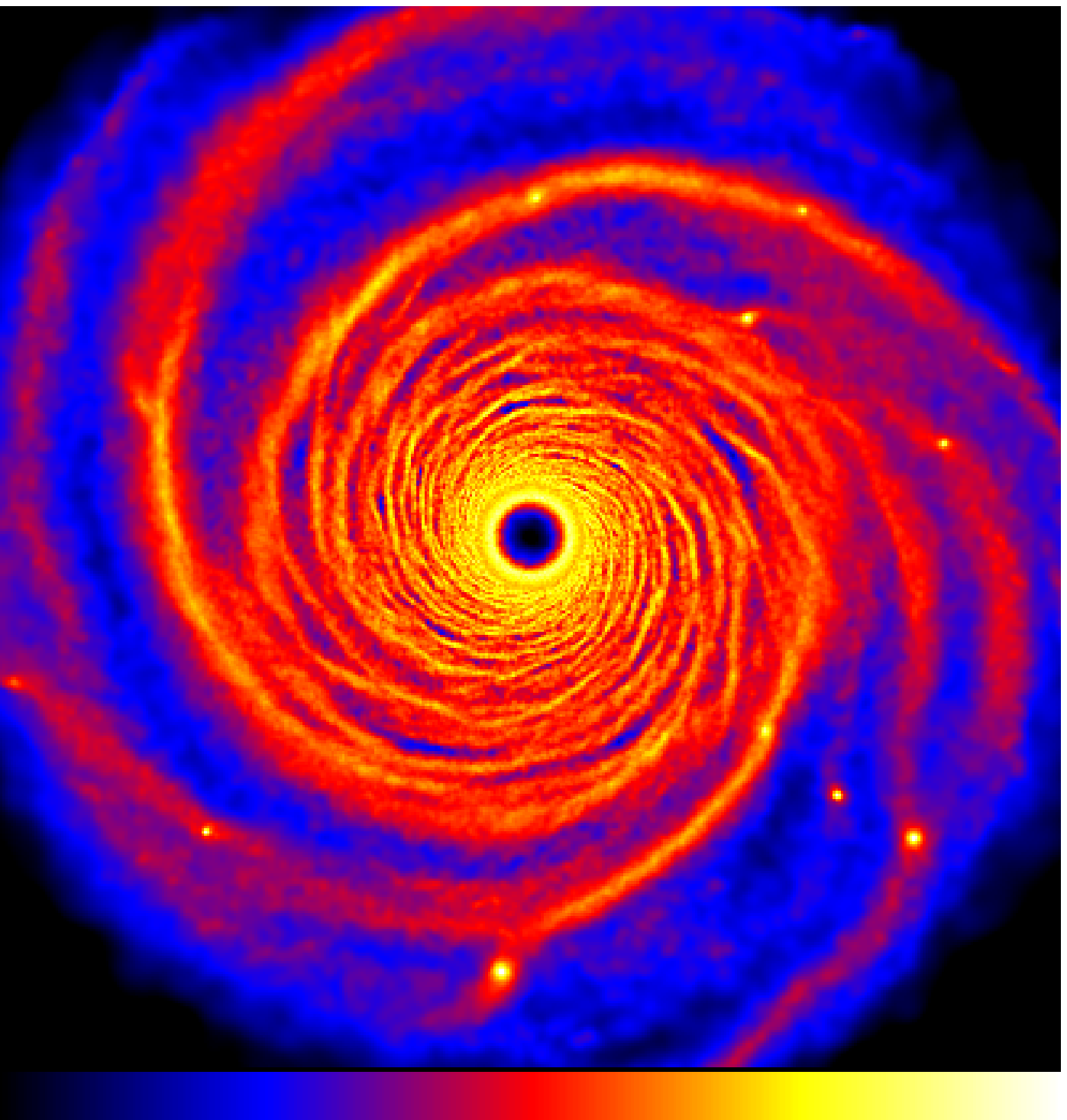,width=40.0mm}}
\centerline{\epsfig{figure=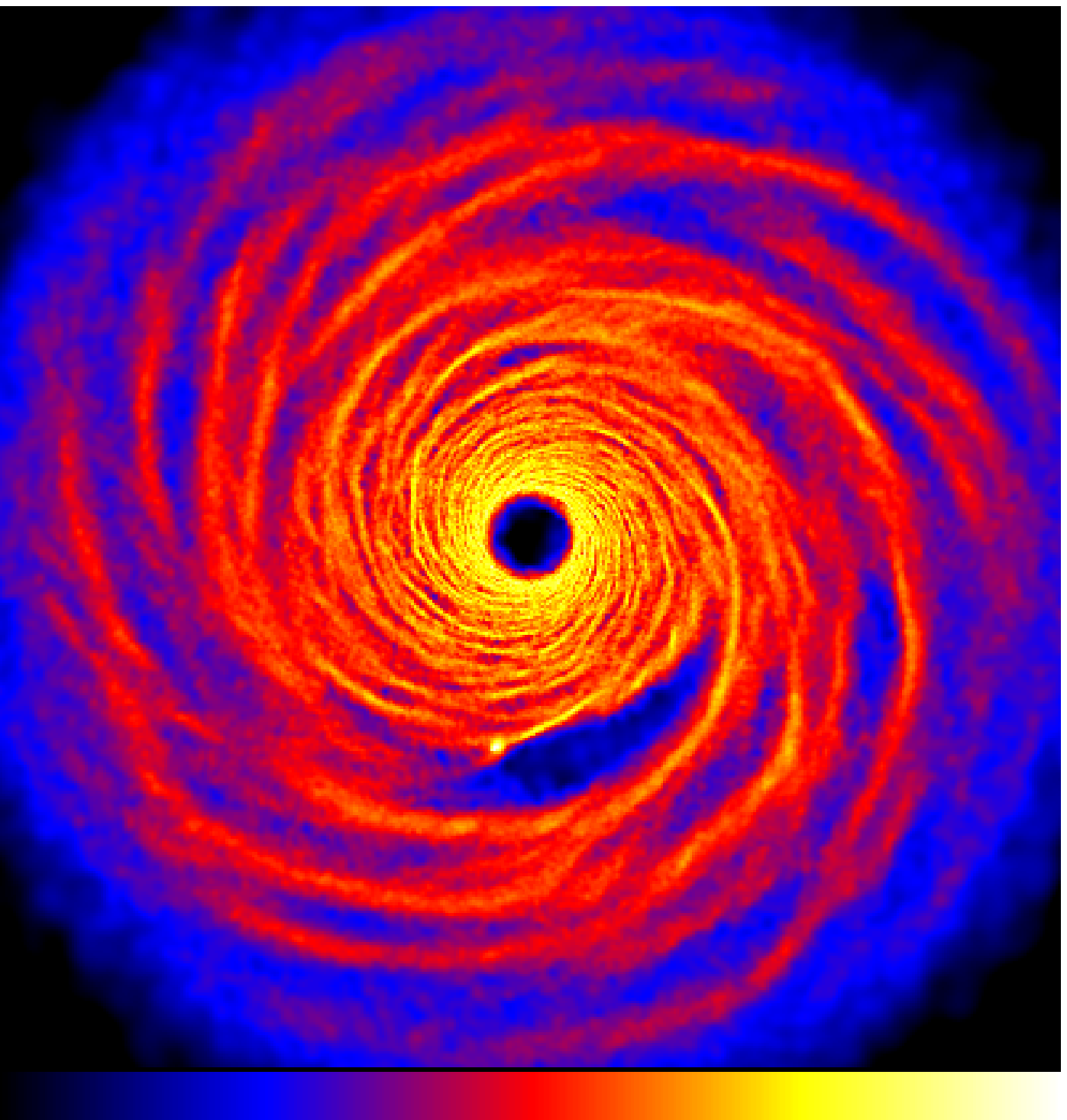,width=40.0mm}
            \epsfig{figure=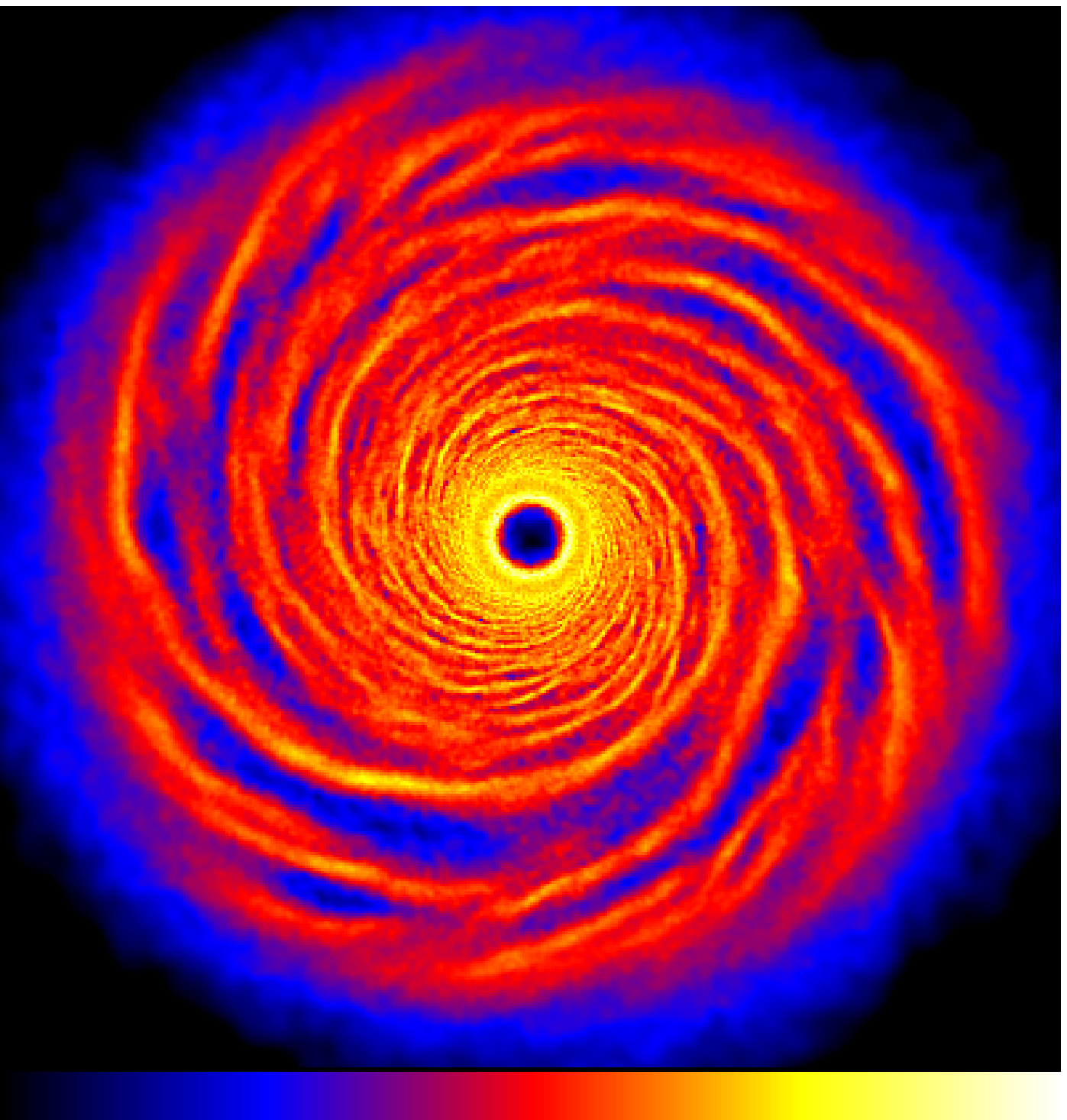,width=40.0mm}}
\caption{Surface density structure of discs with masses $M_{\rm disc}
  = 0.1$ and with cooling times of $t_{\rm cool} \Omega = 3$ (top
  left), $t_{\rm cool} \Omega = 5$ (top right), $t_{\rm cool} \Omega =
  6$ (bottom left), and $t_{\rm cool} \Omega = 7$ (bottom right). The
  logarithmic colour scale in each figure is from $10$ g cm$^{-2}$ to
  $2 \times 10^4$ g cm$^{-2}$. The linear scale is from -25 to 25 for
  both axes.}
\label{Md01_167disk}
\end{figure}

Figure \ref{Md01_167disk} illustrates the procedure discussed above.
The disc mass in all four figures is $M_{\rm disc} = 0.1$, the
specific heat ratio is $\gamma = 5/3$, and the cooling times are
$t_{\rm cool} \Omega = 3$ (top left), $t_{\rm cool} \Omega = 5$ (top
right), $t_{\rm cool} \Omega = 6$ (bottom left), and $t_{\rm cool}
\Omega = 7$ (bottom right). In this particular case we did not
complete a $t_{\rm cool} \Omega = 4$ run since the $t_{\rm cool}
\Omega = 5$ run had already shown signs of fragmentation prior to the
completion of the $t_{\rm cool} \Omega = 4$ simulation. For $t_{\rm
  cool} \Omega = 3$ there are a large number of fragments, consistent
with \citet{rice03b}. For $t_{\rm cool} \Omega = 5$ there are a number
of fragments, while for $t_{\rm cool} \Omega = 6$ there is only a
single fragment that in the image can be seen just below the central
object. For $t_{\rm cool} \Omega = 7$, which ran almost $7$ outer
rotation periods, the disc is clearly unstable at all radii, but there
is no signs of fragmentation.

\begin{table}
\begin{tabular}{|r|r|r|r|r|}
\hline
$M_{\rm disc}/M_*$&$\gamma$&$t_{\rm cool}\Omega$&$E_{\rm tot}$\\
\hline
0.1&5/3&3&$-9.7 \times 10^{-7}$\\
0.1&5/3&5&$-1.0 \times 10^{-7}$\\
0.1&5/3&6&$-3.8 \times 10^{-5}$\\
0.1&5/3&7&no clumps\\
0.1&7/5&11&$-8.8 \times 10^{-7}$\\
0.1&7/5&12&$-6.6 \times 10^{-8}$\\
0.1&7/5&13&no clumps\\
0.25&5/3&5&$-9.4 \times 10^{-6}$\\
0.25&5/3&6&$-3.0 \times 10^{-7}$\\
0.25&5/3&7&no clumps\\
0.25&7/5&11&$-8.2 \times 10^{-7}$\\
0.25&7/5&12&$-7.2 \times 10^{-7}$\\
0.25&7/5&13&no clumps\\
0.5&5/3&6&$-4.9 \times 10^{-5}$\\
0.5&5/3&7&no clumps\\
0.5&7/5&11&$-1.0 \times 10^{-5}$\\
0.5&7/5&12&$-7.5 \times 10^{-6}$\\
0.5&7/5&13&no clumps\\
\hline
\end{tabular}
\caption{Table showing the results of a series of simulations considering
discs with masses between $M_{\rm disc} = 0.1$ and $M_{\rm disc} = 0.5$,
specific heat ratio of $\gamma = 5/3$ and $\gamma = 7/5$, and various cooling
times. These results suggest that the fragmentation boundary does not depend
on disc mass, and that for $\gamma = 7/5$ fragmentation may occur for cooling
times almost twice the local dynamical time.}
\label{tab:res}
\end{table}

We repeated the above procedure for disc masses of $M_{\rm disc} =
0.25$, and $M_{\rm disc} = 0.5$ and for specific heat ratios of $\gamma
= 5/3$ and $\gamma = 7/5$. The results are summarised in Table $1$. The
columns in Table 1 are the ratio of disc to central object mass,
$M_{\rm disc}/M_*$, the specific heat ratio, $\gamma$, the cooling
time, $t_{\rm cool} \Omega$, and if fragmentation occurs, the total
energy (in code units) of the densest clump, $E_{\rm tot}$, where
$E_{\rm tot}$ is the sum of the thermal energy and gravitational
potential energy \citep{bate95}.  In the earlier work of
\citet{rice03c} there was a suggestion that the cooling time required
for fragmentation may depend on the total disc mass, relative to the
mass of the central object.  The results shown in Table 1 suggest that
there is no disc mass dependence. Fragmentation occurs for $t_{\rm
cool} \Omega$ between $6$ and $7$ when $\gamma = 5/3$ and between $12$
and $13$ when $\gamma = 7/5$, for all disc masses considered. The
reason why there is a difference between \citet{rice03c} is
unclear. Their discs had slightly steeper surface density profiles
($\Sigma \propto r^{-7/4}$ rather than $\Sigma \propto r^{-1}$), and it
is possible that their $t_{\rm cool} \Omega = 5$ simulation that did
not fragment, may have done so had it been run for longer. Table 1 also
shows that in all the cases where clumps were detected, the total
energy of the densest clump is negative and that at least the densest
clump is bound.

\begin{figure}
\centerline{\epsfig{figure=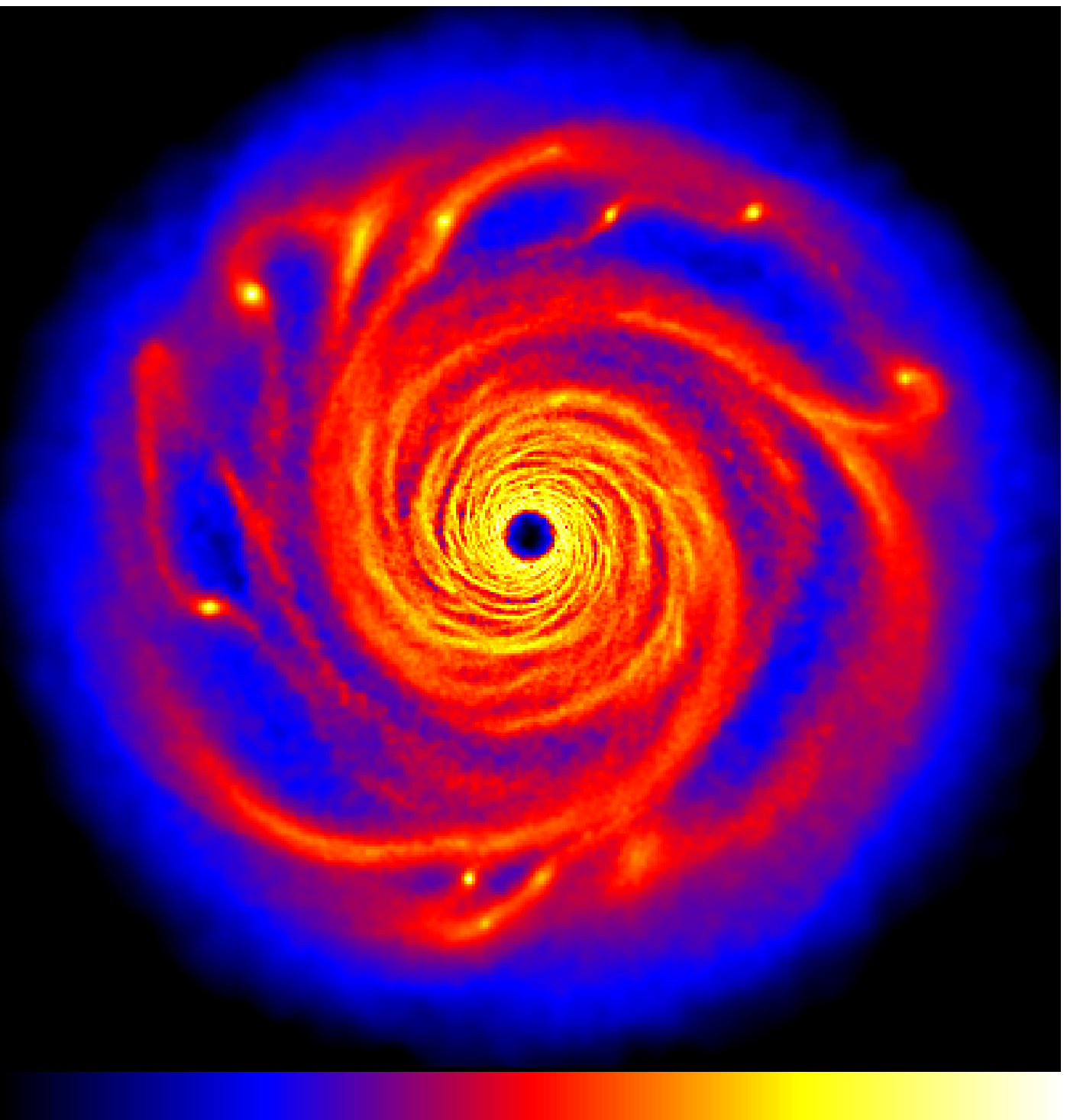,width=40.0mm}
            \epsfig{figure=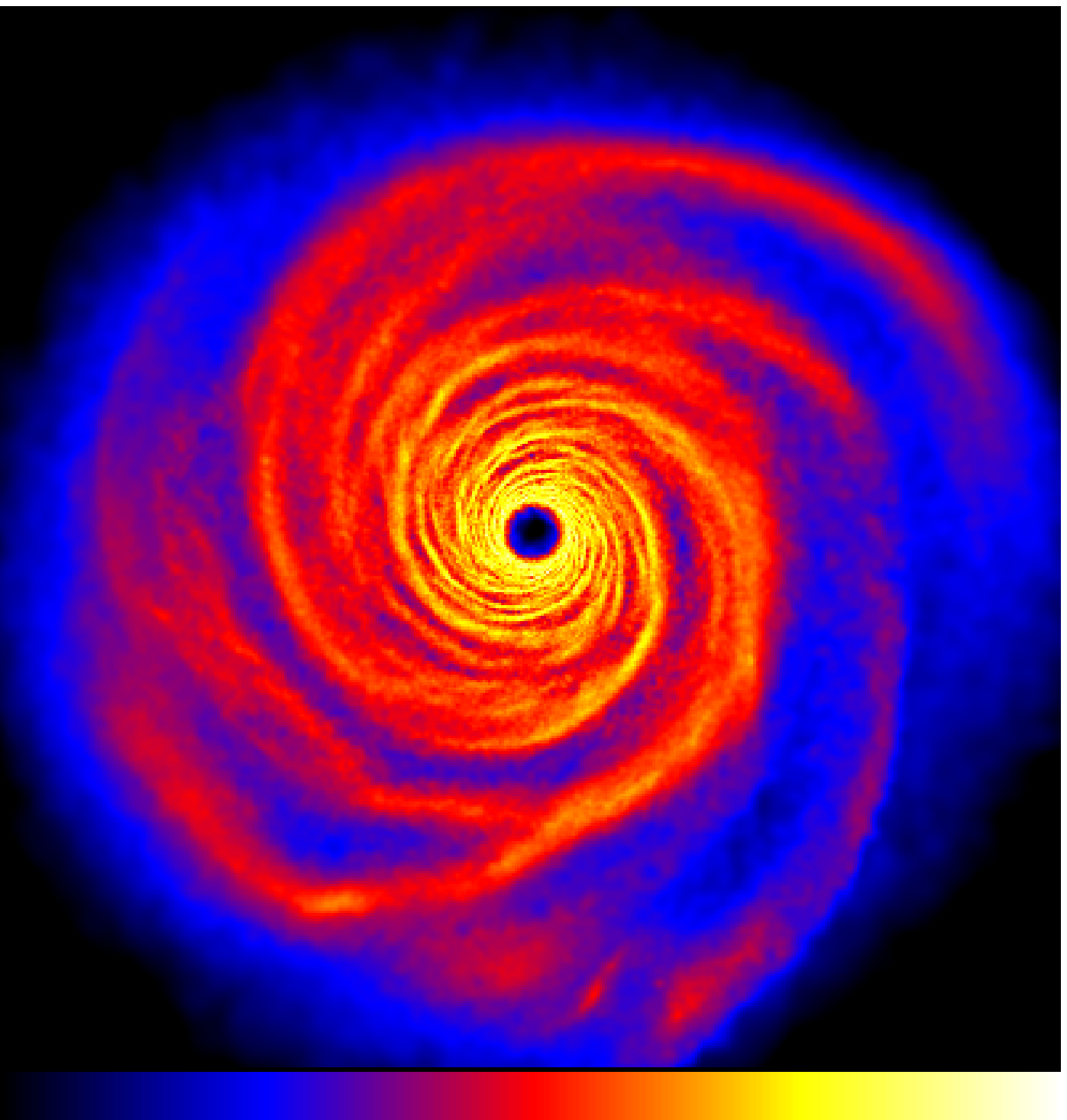,width=40.0mm}}
\caption{Surface density structure of discs with a mass of $M_{\rm
    disc} = 0.25$, a specific heat ratio of $\gamma = 5/3$, and
  cooling times of $t_{\rm cool} \Omega = 6$ (left hand panel) and
  $t_{\rm cool} \Omega = 7$ (right hand panel).  The lack of
  fragmentation in the right hand panel suggests that the
  fragmentation boundary is at a cooling time of between $t_{\rm cool}
  \Omega = 6$ and $t_{\rm cool} \Omega = 7$. The colour scale of the
  density and the linear scale of the image are the same as in
  Fig. 1. }
\label{Md025_167disk}
\end{figure}

\begin{figure}
\centerline{\epsfig{figure=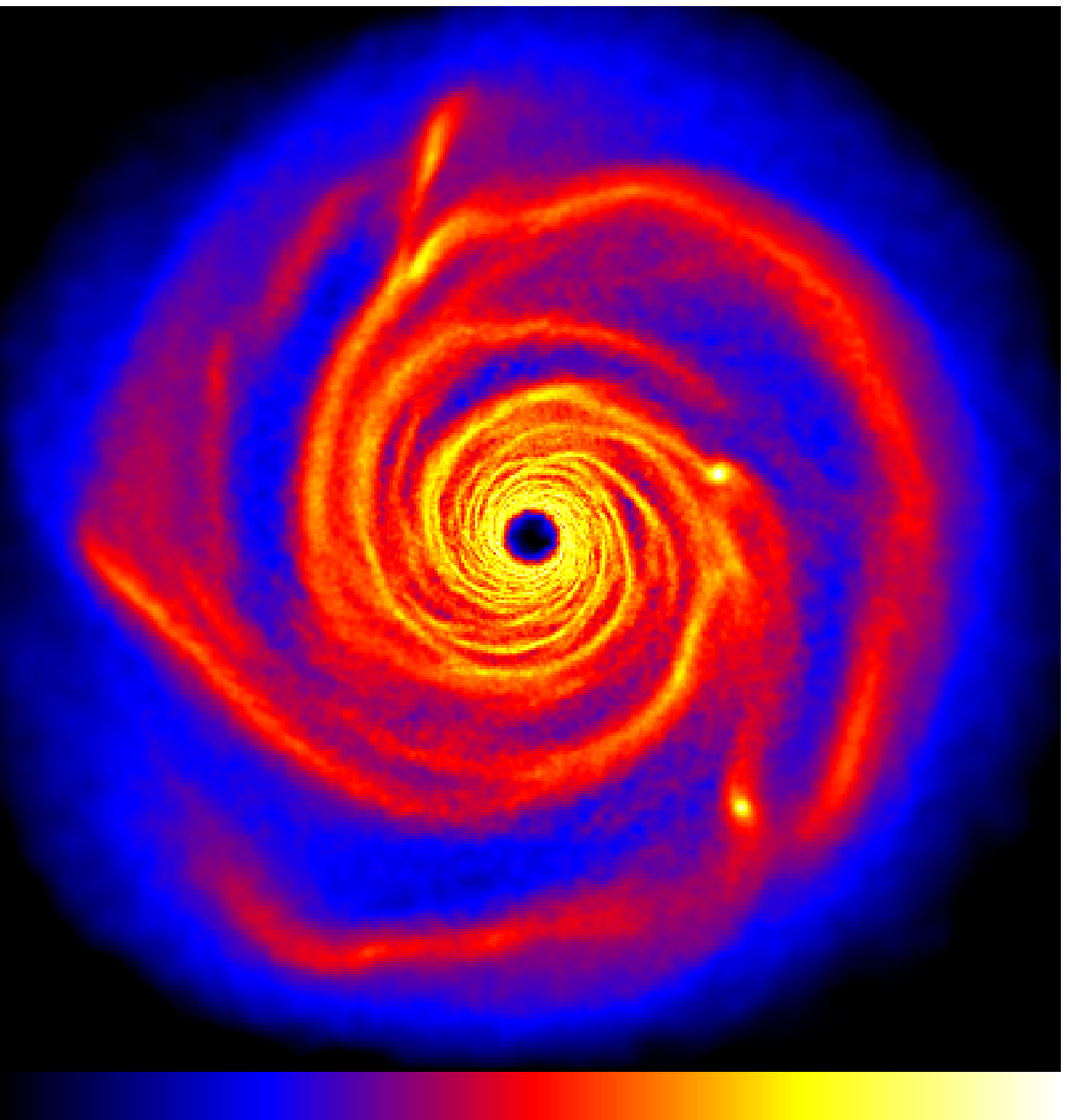,width=40.0mm}
            \epsfig{figure=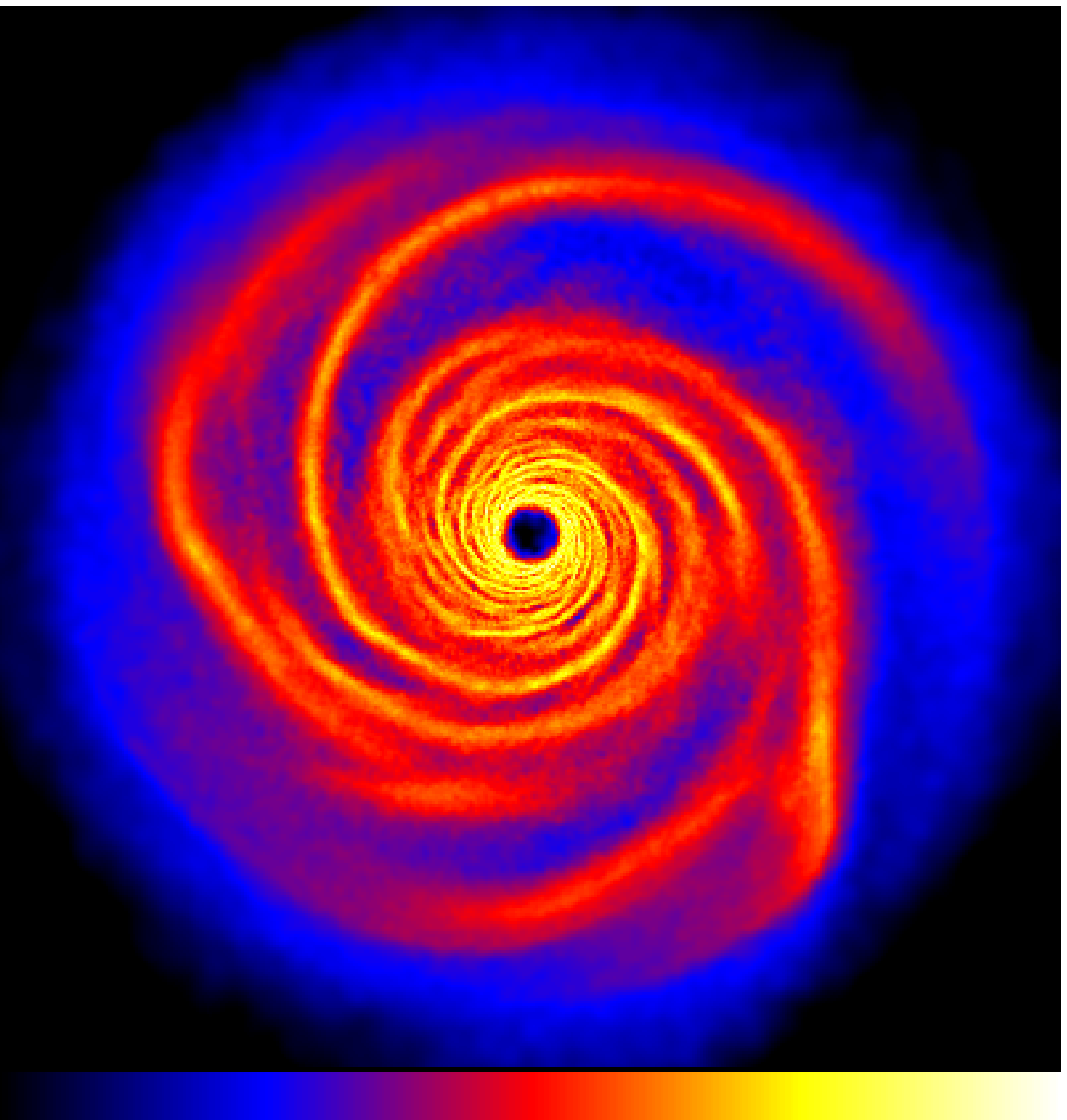,width=40.0mm}}
\caption{Surface density structure of discs with a the same mass as in
  Figure \ref{Md025_167disk} \, but with a specific heat ratio of
  $\gamma = 7/5$, and cooling times of $t_{\rm cool} \Omega = 12$
  (left hand panel) and $t_{\rm cool} \Omega = 13$ (right hand panel).
  The lack of fragmentation in the right hand panel in this case
  suggests that for $\gamma = 7/5$ the fragmentation boundary is at a
  cooling time of between $t_{\rm cool} \Omega = 12$ and $t_{\rm cool}
  \Omega = 13$.  The colour scale of the density and the linear scale
  of the image are the same as in Fig. 1. }
\label{Md025_14disk}
\end{figure}

Although Table 1 shows that the fragmentation boundary occurs for
cooling times longer than that predicted by \citet{gammie01}, for
$\gamma = 5/3$ the required cooling time is still smaller than the
local dynamical time. It also shows that as the specific heat ratio
decreases, the required cooling time increases and is almost twice the
local dynamical time for $\gamma = 7/5$.  The fragmentation boundary
for a disc mass of $M_{\rm disc} = 0.25$ and for both of the specific
heat ratios considered is shown in Figures \ref{Md025_167disk} and
\ref{Md025_14disk}.  Figure \ref{Md025_167disk} shows the final
surface density structures for $M_{\rm disc} = 0.25$, a specific heat
ratio of $\gamma = 5/3$, and for cooling times of $t_{\rm cool} \Omega
= 6$ (left panel) and $t_{\rm cool} \Omega = 7$ (right panel). The
$t_{\rm cool} \Omega = 7$ simulation was evolved for almost an outer
rotation period longer than the $t_{\rm cool} \Omega = 6$ simulation
yet shows no signs of fragmentation. The discs shown in Figure
\ref{Md025_14disk} have the same parameters as in Figure
\ref{Md025_167disk} except $\gamma = 7/5$, and the cooling times are
$t_{\rm cool} \Omega = 12$, and $t_{\rm cool} \Omega = 13$. Again
there is no sign of fragmentation in the right hand panel which was
also evolved for almost an outer rotation period longer than the
simulation shown in the left hand panel.

As a further numerical check, we repeated one set of calculations using
125000 particles rather than 250000 particles. We considered only the 
case where $M_{\rm disc} = 0.25$ and $\gamma = 5/3$. The result with 125000
particles was the 
same as the simulation with 250000  particles.  Fragmentation occured 
for $t_{\rm cool} = 6 \Omega^{-1}$ and did not occur for $t_{\rm cool} = 
7 \Omega^{-1}$.  Therefore, not only do the simulations that fragment 
satisfy the Jeans Criterion for fragmentation \citep{bate97}, it
appears that the results are resolution independent.

\begin{figure}
\centerline{\epsfig{figure=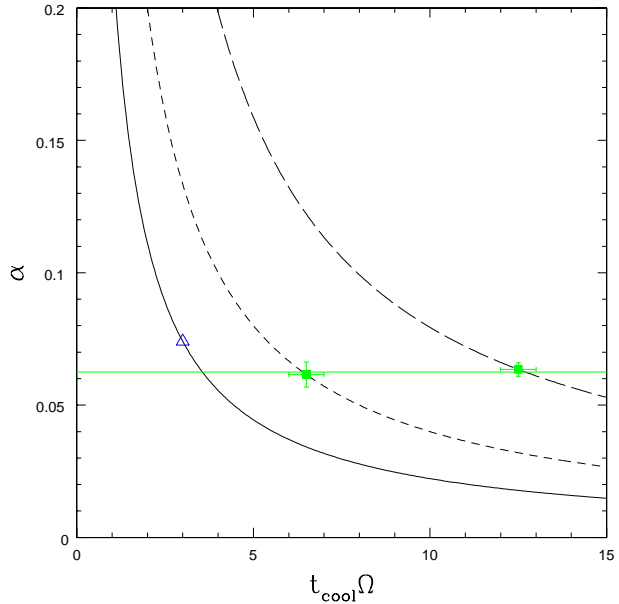,width=84mm}}
\caption{The relationship defined by equation \ref{alpha} for
$\gamma=2$ (solid line), $\gamma=5/3$ (short-dashed line) and
$\gamma=7/5$ (long-dashed line). The data points show the couples
($t_{\rm frag} \Omega,\alpha_{\rm max}$) as derived from the simulations: the
green squares refer to our simulations, while the blue triangle refers to
\citet{gammie01}. The horizontal green line illustrates the constant values
$\alpha=0.06251$.}
\label{fig:alpha}
\end{figure}

\section{A maximum value for gravitational stresses}
\label{disc}

Based on the results summarized in Table \ref{tab:res}, for every value
of $M_{\rm disc}/M_*$ and $\gamma$, we can define a minimum cooling
time for which no fragmentation occurs, $t_{\rm nf}$ and a maximum
cooling time for which fragmentation does occur, $t_{\rm f}$.  The
boundary value of $t_{\rm cool}$ for fragmentation can therefore be
defined as $t_{\rm frag}=1/2(t_{\rm nf}+t_{\rm f})$, with a
corresponding uncertainty given by $\Delta t_{\rm frag}=1/2(t_{\rm nf}
-t_{\rm f})$. The stress $\alpha_{\rm max}$, corresponding to $t_{\rm
frag}$, can be computed from equation (\ref{alpha}), and the
corresponding uncertainty is given by $\Delta\alpha_{\rm
max}=(\alpha_{\rm max}/t_{\rm frag})\Delta t_{\rm frag}$. The resulting
values of $t_{\rm frag}$ and $\alpha_{\rm max}$ are shown as data
points in Fig. \ref{fig:alpha}, together with the curves defined by
equation (\ref{alpha}), for three values of $\gamma=$2, 5/3 and
7/5. The filled green squares with error bars refer to the simulations
presented here. The open blue triangle represent the value found by
\citet{gammie01} in his local, 2D simulations that assumed
$\gamma=2$. This is consistent with our result which suggests that for
$\gamma = 2$, fragmentation should occur between $t_{\rm cool} \Omega =
3$ and $t_{\rm cool} \Omega = 4$. In fact, it is worth noting that
since Gammie's simulations are 2D, we should not expect a perfect
agreement between our 3D results and his ones. This can be partially
seen already from Fig. \ref{fig:alpha}. In particular, care should be
taken in considering the role of the adiabatic index $\gamma$, which
has a different physical interpretation in 2D and in 3D. However, as
discussed in more detail in \citet{gammie01}, a mapping is possible
between the 2D and the 3D adiabatic indeces. In the case of
self-gravitating discs, Gammie's choice of a 2D adiabatic index equal
to 2 does correspond to $\gamma=2$ also in 3D \citep{gammie01}.

As can be seen, fragmentation occurs at an almost constant value of
$\alpha$ ($\alpha_{\rm max} \sim 0.06$, indicated by the horizontal
green line in Fig. \ref{fig:alpha}), thus vindicating the idea that
gravitational instabilities cannot provide (in a steady state) a stress
larger than $\alpha_{\rm max}$. If the dissipation associated with
$\alpha_{\rm max}$ is not sufficient to balance the cooling rate, then
the reaction of the disc is to fragment into bound objects.

\section{Discussion and conclusions}

In this paper we elucidate the processes that lead to the fragmentation
of a massive disc.  Our main result is the determination of a maximum
value for the stress that can be provided by gravitational
instabilities in a quasi-steady state. We then argue that fragmentation
will occur whenever the external cooling requires, in order to be
balanced by internal heating, a stress larger than this maximum value,
that we estimate to be $\alpha_{\rm max}\sim 0.06$ (in units of the
local disc pressure). As a consequence, discs with larger values of the
ratio of the specific heats will be less susceptible to
fragmentation. For $\gamma=7/5$, for example, we estimate the
fragmentation cooling time to be between $11\Omega^{-1}$ and
$12\Omega^{-1}$, compared to between $3 \Omega^{-1}$ and $4
\Omega^{-1}$ for $\gamma = 2$ \citep{gammie01}.

We wish to stress that the threshold value for $\alpha$ that we have
found here refers to a quasi-steady state, in which the disc stays in
thermal equilibrium and the relevant physical quantities do not change
significantly on time scales shorter than the thermal timescale. We
have already shown \citep{LR05} how very massive discs (with masses
comparable to that of the central object) can generate transient strong
spiral episodes, with correspondingly large values of the stress
$\alpha$, which, however, do not last for longer than one dynamical
timescale (see details in \citealt{LR05}). 

A further remark is in order, in reference to the possibility of
non-local transport in self-gravitating discs. In all our simulations,
we did not find any significant evidence for non-local transport of
energy due to self-gravity \citep{LR04,LR05}. If the disc does not
fragment, the dissipation provided by the gravitational stresses
balances almost exactly the imposed cooling rate. However, this
conclusion might depend on the simulation setup and, in particular, on
the assumed radial dependence of the cooling time. \citet{mejia05}
claim to find evidence for non-local energy transport in their
simulations that employ a $t_{\mathrm{cool}}$ constant with radius
(rather that $\propto \Omega^{-1}$, as we do). If this is the case, it
might offer a possible escape route for the disc in order to avoid
fragmentation. Consider the case where $t_{\mathrm{cool}}\Omega$ is
such that the disc is stable against fragmentation in the inner
regions, but would fragment in the outer regions, following our
prediction in equation (\ref{eq:amax}) (which is based on the implicit
assumption that energy dissipation is viscous). In such a situation,
the inner disc could ``help'' the outer disc, by heating it up via
non-local effects and preventing fragmentation. This process is not
viable in the simulations presented here, since here the whole disc is
uniformly stable or unstable with respect to fragmentation and
increasing the cooling rate of the inner disc via non-local energy
transport would make it fragment. Clearly, further simulations of the
fragmentation process in discs with a radius-dependent
$t_{\mathrm{cool}}\Omega$ are needed to investigate this issue.

Finally, we note that the fragmentation requirements determined by
\citet{rafikov05} were calculated assuming the much shorter cooling
times of \citet{gammie01}, and by assuming that the Toomre $Q$
parameter must be unity or less for fragmentation to commence. Since
the actual $Q$ value required for fragmentation can be slightly
greater than $1$ \citep{pickett98}, and since the required cooling
time in discs with $\gamma = 7/5$ (as used by \citet{boss98,boss02})
can be larger than that predicted by \citep{gammie01}, the conditions
for fragmentation may not be as stringent as those determined by
\citet{rafikov05}.

\section*{acknowledgements}
The computations reported here were performed using the UK
Astrophysical Fluids Facility (UKAFF), and Datastar at the San Diego
Supercomputing Center (SDSC). This work was supported by NASA under
grants NAG5-13207 and NNG04GL01G from the Origins of Solar Systems and
Astrophysics Theory Programs, and by the NSF under grant
AST~0407040. WKMR acknowledges support from the Institute for Pure and
Applied Mathematics (IPAM) at the University of California, Los
Angeles, where some of this work was carried out. We thank Cathie
Clarke and Matthew Bate for useful discussions and Jim Pringle for a
careful reading of the manuscript.

\bibliographystyle{mn2e} 

\bibliography{frag}

\begin{thebibliography}{}

\bibitem[\protect\citeauthoryear{Artymowicz \& Lubow}{Artymowicz \&
  Lubow}{1994}]{lubow94}
Artymowicz P.,  Lubow S.~H.,  1994, ApJ, 421, 651

\bibitem[\protect\citeauthoryear{Balbus \& Papaloizou}{Balbus \&
  Papaloizou}{1999}]{balbus99}
Balbus S.~A.,  Papaloizou J. C.~B.,  1999, ApJ, 521, 650

\bibitem[\protect\citeauthoryear{Bate, Bonnell \& Price}{Bate
  et~al.}{1995}]{bate95}
Bate M.~R.,  Bonnell I.~A.,    Price N.~M.,  1995, MNRAS, 277, 362

\bibitem[\protect\citeauthoryear{Bate \& Burkert}{Bate \&
  Burkert}{1997}]{bate97}
Bate M.~R.,  Burkert A.,  1997, MNRAS, 288, 1060

\bibitem[\protect\citeauthoryear{Beltran et~al.,}{Beltran
  et~al.}{2004}]{beltran04}
Beltran M.,  et~al., 2004, ApJ, 601, L187

\bibitem[\protect\citeauthoryear{Benz}{Benz}{1990}]{benz90}
Benz W.,  1990, in Buchler J.,  ed., The Numerical Modeling of Nonlinear
  Stellar Pulsations Kluwer, Dordrecht

\bibitem[\protect\citeauthoryear{Boss}{Boss}{1998}]{boss98}
Boss A.~P.,  1998, Nature, 393, 141

\bibitem[\protect\citeauthoryear{Boss}{Boss}{2002}]{boss02}
Boss A.~P.,  2002, ApJ, 576, 462

\bibitem[\protect\citeauthoryear{Chini et~al.,}{Chini  et~al.}{2004}]{chini04}
Chini R.,  et~al., 2004, Nature, 429, 155

\bibitem[\protect\citeauthoryear{Gammie}{Gammie}{1996}]{gammie96}
Gammie C.~F.,  1996, ApJ, 457, 355

\bibitem[\protect\citeauthoryear{Gammie}{Gammie}{2001}]{gammie01}
Gammie C.~F.,  2001, ApJ, 553, 174

\bibitem[\protect\citeauthoryear{Goodman \& Tan}{Goodman \&
  Tan}{2004}]{goodman04}
Goodman J.,  Tan J.~C.,  2004, ApJ, 608, 108

\bibitem[\protect\citeauthoryear{Greenhill \& Gwinn}{Greenhill \&
  Gwinn}{1997}]{greenhill97}
Greenhill L.~J.,  Gwinn C.~R.,  1997, Astrophysics \& Space Science, 248, 261

\bibitem[\protect\citeauthoryear{Johnson \& Gammie}{Johnson \&
  Gammie}{2003}]{johnson03}
Johnson B.~M.,  Gammie C.~F.,  2003, ApJ, 597, 131

\bibitem[\protect\citeauthoryear{Kondratko, Greenhill \& Moran}{Kondratko
  et~al.}{2005}]{kondratko05}
Kondratko P.~T.,  Greenhill L.~J.,    Moran J.~M.,  2005, ApJ, 618, 618

\bibitem[\protect\citeauthoryear{Kuiper}{Kuiper}{1951}]{kuiper51}
Kuiper G.,  1951, in Hynek J. e.~a.,  ed., Topical symposium commemorating the
  50th anniversary of the Yerkes Observatory and half a century of progess in
  astrophysics On the origin of the solar system.
McGraw Hill, p.~357

\bibitem[\protect\citeauthoryear{Larson}{Larson}{1984}]{larson84}
Larson R.,  1984, MNRAS, 206, 197

\bibitem[\protect\citeauthoryear{Laughlin \& Bodenheimer}{Laughlin \&
  Bodenheimer}{1994}]{laughlin94}
Laughlin G.,  Bodenheimer P.,  1994, ApJ, 436, 335

\bibitem[\protect\citeauthoryear{Lin \& Pringle}{Lin \&
  Pringle}{1987}]{linpringle87}
Lin D. N.~C.,  Pringle J.~E.,  1987, MNRAS, 225, 607

\bibitem[\protect\citeauthoryear{Lodato \& Bertin}{Lodato \&
  Bertin}{2003}]{LB03a}
Lodato G.,  Bertin G.,  2003, A\&A, 398, 517

\bibitem[\protect\citeauthoryear{Lodato \& Rice}{Lodato \& Rice}{2004}]{LR04}
Lodato G.,  Rice W. K.~M.,  2004, MNRAS, 351, 630

\bibitem[\protect\citeauthoryear{Lodato \& Rice}{Lodato \& Rice}{2005}]{LR05}
Lodato G.,  Rice W. K.~M.,  2005, MNRAS, 358, 1489

\bibitem[\protect\citeauthoryear{Matsumoto \& Tajima}{Matsumoto \&
  Tajima}{1995}]{matsumoto95}
Matsumoto R.,  Tajima T.,  1995, ApJ, 445, 767

\bibitem[\protect\citeauthoryear{Mayer, Quinn, Wadsley \& Stadel}{Mayer
  et~al.}{2004}]{mayer04}
Mayer L.,  Quinn T.,  Wadsley J.,    Stadel J.,  2004, ApJ, 609, 1045

\bibitem[\protect\citeauthoryear{Mejia, Durisen, Pickett \& Cai}{Mejia
  et~al.}{2005}]{mejia05}
Mejia A.~C.,  Durisen R.~H.,  Pickett M.~K.,    Cai K.,  2005, ApJ, 619, 1098

\bibitem[\protect\citeauthoryear{Monaghan}{Monaghan}{1992}]{monaghan92}
Monaghan J.~J.,  1992, ARA\&A, 30, 543

\bibitem[\protect\citeauthoryear{Murray}{Murray}{1996}]{murray96}
Murray J.~R.,  1996, MNRAS, 279, 402

\bibitem[\protect\citeauthoryear{Pickett et~al.,}{Pickett
  et~al.}{1998}]{pickett98}
Pickett B.~K.,  et~al., 1998, ApJ, 504, 468

\bibitem[\protect\citeauthoryear{Pringle}{Pringle}{1981}]{pringle81}
Pringle J.~E.,  1981, ARA\&A, 19, 137

\bibitem[\protect\citeauthoryear{Rafikov}{Rafikov}{2005}]{rafikov05}
Rafikov R.,  2005, ApJ, 621, 69

\bibitem[\protect\citeauthoryear{Rice, Armitage, Bate \& Bonnell}{Rice
  et~al.}{2003}]{rice03c}
Rice W. K.~M.,  Armitage P.~J.,  Bate M.~R.,    Bonnell I.~A.,  2003, MNRAS,
  339, 1025

\bibitem[\protect\citeauthoryear{Rice, Armitage, Bate, Bonnell, Jeffers \&
  Vine}{Rice et~al.}{2003}]{rice03b}
Rice W. K.~M.,  Armitage P.~J.,  Bate M.~R.,  Bonnell I.~A.,  Jeffers S.~V.,
  Vine S.~G.,  2003, MNRAS, 346, L36

\bibitem[\protect\citeauthoryear{Rodriguez, Loinard, D'Alessio, Wilner \&
  P.T.P.}{Rodriguez et~al.}{2005}]{rodriguez05}
Rodriguez L.~F.,  Loinard L.,  D'Alessio P.,  Wilner D.,    P.T.P. H.,  2005,
  ApJ, 621, L133

\bibitem[\protect\citeauthoryear{Shakura \& Sunyaev}{Shakura \&
  Sunyaev}{1973}]{shakura73}
Shakura N.~I.,  Sunyaev R.~A.,  1973, A\&A, 24, 337

\end{thebibliography}

\end{document}